\documentclass{amsart}

\newtheorem{theorem}{Theorem}[section]
\newtheorem{lemma}[theorem]{Lemma}
\newtheorem{corollary}[theorem]{Corollary}
\newtheorem{proposition}[theorem]{Proposition}

\theoremstyle{definition}
\newtheorem{definition}[theorem]{Definition}

\theoremstyle{remark}
\newtheorem{remark}[theorem]{Remark}

\newcommand{\cB}{{\mathcal B}}

\newcommand{\cP}{{\mathcal P}}

\newcommand{\cS}{{\mathcal S}}

\newcommand{\cJ}{{\mathfrak{J}}}
\newcommand{\Jm}{J_{\mathrm{m}}}
\newcommand{\bR}{{\mathbb{R}}}
\newcommand{\bC}{{\mathbb{C}}}

\numberwithin{equation}{section}

\newcommand{\Tr}{\mathrm{Tr}}
\newcommand{\id}{\mathrm{id}}
\newcommand{\vr}{\varrho}
\newcommand{\rr}{\vr^{1/2}}

\newcommand{\jed}{{\mathbb{I}}}
\begin{document}

\title[$k$-decomposability]{On $k$-decomposability of positive maps}

\author[L.E. Labuschagne]{Louis E. Labuschagne}
\address{Department of Maths, Applied Maths and Astronomy,
P.O.Box 392, University of South Africa, 0003 Pretoria, South
Africa} \email{labusle@unisa.ac.za}
\author[W.A. Majewski]{W{\l}adys{\l}aw A. Majewski}
\address{Institute of Theoretical Physics and Astrophysics, Gda{\'n}sk
University, Wita Stwo\-sza~57, 80-952 Gda{\'n}sk, Poland} \email{fizwam@univ.gda.pl}
\author[M. Marciniak]{Marcin Marciniak}
\address{Institute of Mathematics, Gda{\'n}sk University,
Wita Stwosza 57, 80-952 Gda{\'n}sk, Po\-land}
\email{matmm@univ.gda.pl}
\thanks{L.E.L. and W.A.M. are supported by Poland-South Africa Cooperation Joint Project 
while M.M. is supported by KBN grant 2P03A00723
}

\begin{abstract}
We extend the theory of decomposable maps by giving a detailed description
of $k$-positive maps. A relation between transposition and modular theory is
established. The structure of positive maps in terms of modular theory 
(the generalized Tomita-Takesaki scheme) is examined.
\end{abstract}

\maketitle

\section{Definitions, notations and stating the problem}
For any $C^*$-algebra $A$ let $A^+$ denote the set of all positive
elements in $A$. A~{\it state} on a~unital $C^*$-algebra $A$ is a
linear functional $\omega:A\to \mathbb{C}$ such that
$\omega(a)\geq 0$ for every $a\in A^+$ and $\omega(\mathbb{I})=1$
where $\mathbb{I}$ is the unit of $A$. By $\cS(A)$ we will denote
the set of all states on $A$. For any Hilbert space $H$ we denote
by $\cB(H)$ the set of all bounded linear operators on $H$.

A linear map $\varphi:A\to B$ between $C^*$-algebras is called {\it positive}
if $\varphi(A^+)\subset B^+$. For $k\in\mathbb{N}$ we consider a~map
$\varphi_k:M_k(A)\to M_k(B)$ where $M_k(A)$ and $M_k(B)$ are the algebras
of $k\times k$
matrices with coefficients from $A$ and $B$ respectively, and
$\varphi_k([a_{ij}])=[\varphi(a_{ij})]$.
We say that $\varphi$ is {\it $k$-positive}
if the map $\varphi_k$ is positive. The map $\varphi$ is said to be
{\it completely positive}
when it is $k$-positive for every $k\in\mathbb{N}$.

A {\it Jordan morphism} between $C^*$-algebras $A$ and $B$ is
a~linear map ${\cJ}:A\to B$ which respects the Jordan structures
of algebras $A$ and $B$, i.e. $\cJ(ab+ba)= \cJ(a)\cJ(b)+
\cJ(b)\cJ(a)$ for every $a,b\in A$. Let us recall that every
Jordan morphism is a positive map but it need not be a completely
positive one (in fact it need not even be 2-po\-si\-ti\-ve). It is
commonly known (\cite{Sto3}) that every Jordan morphism $\cJ:A\to
\cB(H)$ is a sum of a $^*$-morphism and a $^*$-antimorphism.

The Stinespring theorem states that every completely positive map
$\varphi:A\to \cB(H)$ has the form $\varphi(a)=W^*\pi(a)W$, where
$\pi$ is a $^*$-representation of $A$ on some Hilbert space $K$,
and $W$ is a bounded operator from $H$ to $K$.

Following St{\o}rmer (\cite{Sto1}) we say that a map
$\varphi:A\to \cB(H)$ is {\it decomposable} if there are a Hilbert space $K$,
a Jordan morphism $\cJ:A\to \cB(K)$, and a bounded linear operator
$W$ from $H$ to $K$ such that $\varphi(a)=W^*\cJ(a)W$ for every $a\in A$.

Let $(e_i)$ be a fixed orthonormal basis in some Hilbert space $H$.
Define a conjugation $J_c$ associated with this basis by the formula
$J_c\left(\sum\limits_i\lambda_ie_i\right)=\sum\limits_i\overline{\lambda_i}e_i$.
The map $J_c$ has the following properties:
(i)~$J_c$ is an antilinear isomorphism of $H$;
(ii)~$J_c^2=\mathbb{I}$;
(iii)~$\langle J_c\xi, J_c\eta\rangle=\langle\eta,\xi\rangle$ for every $\xi,\eta\in H$;
(iv)~the map $a\mapsto J_caJ_c$ is a $^*$-automorphism of the
algebra $\cB(H)$.
For every $a\in \cB(H)$ we denote by $a^t$ the element $J_ca^*J_c$
and we call it a {\em transposition} of the element $a$. From the
above properties (i) -- (iv) it follows that the
\textit{transposition map} $a\mapsto a^{\mathrm{t}}$ is a linear
$^*$-antiautomorphism of $\cB(H)$.

We say that a linear map $\varphi:A\to \cB(H)$ is
\textit{$k$-copositive} (resp. \textit{completely copositive}) if
the map $a\mapsto \varphi(a)^{\mathrm{t}}$ is $k$-positive (resp.
\textit{completely positive}). The following theorem characterizes
decomposable maps in the spirit of Stinespring's theorem:
\begin{theorem}[\cite{Sto2}]
Let $\varphi:A\to \cB(H)$ be a linear map. Then the fo\-llo\-wing conditions
are equivalent:
\begin{itemize}
\item[(i)]
$\varphi$ is decomposable;
\item[(ii)]
for every natural number $k$ and for every matrix
$[a_{ij}]
\in M_k(A)$
such that both $[a_{ij}]$ and $[a_{ji}]$ belong to $M_k(A)^+$ the matrix $[\varphi(a_{ij})]$
is in $M_k(\cB(H))^+$;
\item[(iii)]
there are maps $\varphi_1,\varphi_2:A\to \cB(H)$ such that
$\varphi_1$ is completely positive and $\varphi_2$ completely
copositive, with $\varphi=\varphi_1+\varphi_2$.
\end{itemize}
\end{theorem}

In spite of the enormous efforts,
the classification of decomposable maps is still not
complete even in the case when $A$ and $H$ are finite dimensional, i.e.
$A=\cB(\bC^m)$ and $H=\mathbb{C}^n$.
The most important step was done by St{\o}rmer \cite{Sto2}, Choi \cite{Ch2,Ch3} and
Woronowicz \cite{Wor}.
St{\o}rmer and Woronowicz proved that if $m=n=2$ or $m=2$, $n=3$
then every positive map is decomposable.
The first examples of nondecomposable maps was given by Choi (in the case $m=n=3$)
and Woronowicz (in the case $m=2$, $n=4$).
It seems that the main difficulty in carrying out the
classification of positive maps is the question of the canonical
form of non-decomposable maps. As far as we know there are only
special examples of maps from that class which are scattered
across the literature \cite{Wor,Ch3,KK,Ha,EK,Kos,Rob1,Sto5}. In fact
it seems that in the infinite dimensional case all known examples
of non-decomposable maps rely on deep structure theory of the
underlying algebras. (See for example \cite{Sto5}.) On the other
hand, it seems that very general positive maps (so not of the CP
class) and hence possibly non-decomposable ones, are crucial for
an analysis of nontrivial quantum correlations, i.e. for an
analysis of genuine quantum maps \cite{Wit,Per, Hor, MajO, MajL}.
Having that motivation in mind we wish to present a step toward a
canonical prescription for the construction of decomposable and
non-decomposable maps. Namely, we study the notion of
$k$-decomposability and prove an analog of Theorem 1.1. The basic
strategy of the paper is to employ two dual pictures: one given in
terms of operator algebras while the second one will use the space
of states. Thus, it can be said that we are using the equivalence
of the Schr\"odinger and Heisenberg pictures in the sense of
Kadison \cite{Kad}, Connes \cite{Conn} and Alfsen, Shultz \cite{Alf}.

The paper is organized as follows. In section 2 we recall the
techniques used in \cite{MM} and compare it with results from
\cite{Maj}. In section 3 we formulate our main result concerning
the notion of $k$-decomposability. Section 4 is devoted to a
modification of Tomita-Takesaki theory. Section 5, based on the
previous Section, presents a description of k-decomposibility at
the Hilbert space level. Section 6 provides new results on partial
transposition which are used to complete the description of
k-decomposability.

\section{Dual construction}
Let us recall the construction of Choi \cite{Ch2} (see also
\cite{MM}) which establishes a one-to-one correspondence between
elements of $\cB(\bC^m)\otimes \cB(\bC^n)$ and linear maps from $\cB(\bC^m)$ to $\cB(\bC^n)$.
Fix some orthonormal basis $e_1,e_2,\ldots,e_m$ (resp.
$f_1,f_2,\ldots,f_n$) in $\mathbb{C}^m$ (resp. $\mathbb{C}^n$) and
by $E_{ij}$ (resp. $F_{kl}$) denote the the matrix units in $\cB(\bC^m)$
(resp. $\cB(\bC^n)$). For any $x\in\mathbb{C}^m$ define the linear
operator $V_x:\mathbb{C}^n\to \mathbb{C}^m\otimes\mathbb{C}^n$ by
$V_xy=x\otimes y$ where $y\in\mathbb{C}^n$. For simplicity, we
write $V_i$ instead of $V_{e_i}$ for every $i=1,\ldots,m$. Observe
that for any $h\in \cB(\bC^m)\otimes \cB(\bC^n)$ we have
\begin{equation}\label{d1}
h=\sum_{i,j=1}^{m}E_{ij}\otimes V_i^*hV_j.
\end{equation}
Consequently, for every $h$ one can define the
map $\varphi_h:\cB(\bC^m)\to \cB(\bC^n)$ by
$$
\varphi_h(E_{ij})=V_i^*hV_j,\;\;\;\;i,j=1,2,\ldots,m.
$$
On the other hand following (\ref{d1}) given a linear map $\varphi:\cB(\bC^m)\to \cB(\bC^n)$
one can reconstruct $h$ by the formula
\begin{equation}\label{d3}
h=\sum_{i,j=1}^{m}E_{ij}\otimes \varphi(E_{ij})=(\id\otimes \varphi)
\left(\sum_{i,j=1}^{m}E_{ij}\otimes E_{ij}\right).
\end{equation}
The main properties of this correspondence we summarize in the following
\begin{theorem}[\cite{Ch2,MM}]\label{dth1}
Let $h^*=h$. Then:
\begin{enumerate}
\item[(i)]
The map $\varphi_h$ is completely positive if and only if $h$ is a positive operator, i.e.
$$
\langle z,hz\rangle\geq 0
$$
for every $z\in \mathbb{C}^m\otimes \mathbb{C}^n$;
\item[(ii)]
The map $\varphi_h$ is positive if and only if
\begin{equation}\label{d5}
\langle x\otimes y, h(x\otimes y)\rangle\geq 0
\end{equation}
for every $x\in\mathbb{C}^m$ and $y\in\mathbb{C}^n$.
\item[(iii)]
The map $\varphi_h$ is decomposable if and only if $\omega(h)\geq 0$ for each state
$\omega$ on $\cB(\bC^m)\otimes \cB(\bC^n)$
such that $\omega\circ (t\otimes\id)$ is also a state.
\end{enumerate}
\end{theorem}
If the operator $h$ fulfills the property (\ref{d5}) we will call it a {\it block-positive}
operator.

In this section we compare Theorem \ref{dth1} with the results
presented in \cite{Maj}. For the reader's convenience we recall
the main theorem from this paper.
\begin{theorem}\label{dth2}
A linear map $\varphi:\cB(\bC^m)\to \cB(\bC^n)$ is positive if and only if
it is of the form
$$
\varphi(a)=\sum_{k,l=1}^n \Tr (ag_{lk})F_{kl},\;\;\;\;a\in \cB(\bC^m)
$$
where $g_{kl}\in \cB(\bC^m)$, $k,l=1,\ldots,n$, satisfy the following condition:
for every $x\in\mathbb{C}^m$ and $\lambda_1,\ldots,\lambda_n\in\mathbb{C}$
\begin{equation}\label{d7}
\sum_{k,l=1}^n\lambda_k\overline{\lambda_l}\langle x,g_{kl}x\rangle\geq 0.
\end{equation}
\end{theorem}
In fact the condition (\ref{d7}) coincides with (\ref{d5}).
\begin{proposition}
Let $A\in \cB(\bC^m)\otimes \cB(\bC^n)$. Then the following conditions are equivalent:
\begin{enumerate}
\item[(i)]
for every $x\in\mathbb{C}^m$ and
$y\in\mathbb{C}^n$
$$
\langle x\otimes y, Ax\otimes y\rangle\geq 0;
$$
\item[(ii)]
for every $x\in\mathbb{C}^m$ and $\lambda_1,\ldots,\lambda_n\in\mathbb{C}$
$$
\sum\limits_{k,l=1}^n\lambda_k\overline{\lambda_l}\langle x,A_{kl}x\rangle\geq 0
$$
where $A_{kl}$ are unique elements
of $\cB(\bC^m)$ such that $A=\sum\limits_{k,l}A_{kl}\otimes F_{kl}$;
\item[(iii)]
for every
$y\in\mathbb{C}^n$ and $\mu_1,\ldots,\mu_m\in\mathbb{C}$
$$
\sum\limits_{i,j=1}^m\mu_i\overline{\mu_j}\langle y,A_{ij}'y\rangle\geq 0
$$
where $A_{ij}'$ are unique elements
of $\cB(\bC^n)$ such that $A=\sum\limits_{i,j}E_{ij}\otimes A_{ij}'$.
\end{enumerate}
\end{proposition}
\begin{proof}
${\it (i)}\Longleftrightarrow {\it (ii)}$ Let the $\lambda$'s be coefficients of the expansion
of $y$ in the basis $\{f_k\}$, i.e. $y=\sum_s\lambda_sf_s$. Then we have
\begin{eqnarray*}
\lefteqn{\langle x\otimes y, A x\otimes y\rangle =
 \sum_{s,t}\lambda_s\overline{\lambda_t}\,\langle x\otimes f_t,A x\otimes f_s\rangle=} \\
&=& \sum_{s,t}\sum_{k,l}\lambda_s\overline{\lambda_t}\,\langle x,A_{kl}x\rangle\,
    \langle f_t,F_{kl}f_s\rangle
= \sum_{k,l}\lambda_k\overline{\lambda_l}\,\langle x,A_{kl}x\rangle.
\end{eqnarray*}
This proves the equivalence.

${\it (i)}\Longleftrightarrow {\it (iii)}$
This follows by the same method.
\end{proof}

The next proposition establishes the connection between the two
constructions
\begin{proposition}
Let $\varphi:\cB(\bC^m)\to \cB(\bC^n)$ be a linear map. If
$$
g=\sum_{k,l}g_{kl}\otimes F_{kl}
$$
where $\{g_{kl}\}$ are operators described in Theorem 2.2 and $h$ is the operator
defined in (\ref{d3}) then $h=g^t$.
\end{proposition}
\begin{proof}
Define the sesquilinear form $(\cdot,\cdot)$ on $\cB(\bC^m)$ by
$(a,b)=\Tr(a^*b)$ for $a,b\in \cB(\bC^m)$. Then $\cB(\bC^m)$ becomes a Hilbert
space and $\{E_{ij}\}$ forms an orthonormal basis. From the
definitions of $h$ and $g$ we get
\begin{eqnarray*}
\lefteqn{h=\sum_{i,j}E_{ij}\otimes \varphi(E_{ij})= \sum_{i,j}\sum_{k,l}\Tr(E_{ij}B_{lk})E_{ij}
\otimes F_{kl}}\\
&=& \sum_{k,l}\left(\sum_{i,j}(E_{ji},g_{lk})E_{ji}\right)^T\otimes F_{kl}=
\sum_{kl}g_{lk}^T\otimes F_{lk}^t=g^t
\end{eqnarray*}
\end{proof}

\section{$k$-decomposability}\label{kdec}
The following theorem characterizes $k$-positivity of a map
$\varphi$ in terms of the properties of the operators $g$ and $h$
and constitutes a generalization of Theorems \ref{dth1} and
\ref{dth2}.
\begin{theorem}
Let $\varphi:\cB(\bC^m)\to \cB(\bC^n)$ be a linear map. Then the following conditions are
equivalent:
\begin{enumerate}
\item[(i)]
$\varphi$ is $k$-positive;
\item[(ii)]
for every $y_1,\ldots,y_m\in \mathbb{C}^n$ such that $\dim{\rm span}\{y_1,\ldots,y_m\}\leq k$
we have
$$
\sum_{i,j=1}^n\langle y_j,h_{ij}y_i\rangle\geq 0
$$
where $h_{ij}\in \cB(\bC^n)$ are such that $h=\sum_{i,j}E_{ij}\otimes h_{ij}$, i.e.
$h_{ij}=\varphi(E_{ij})$;
\item[(iii)]
for every $x_1,\ldots,x_n\in\mathbb{C}^m$ such that $\dim{\rm span}\{x_1,\ldots,x_n\}\leq k$
we have
$$
\sum_{k,l=1}^n\langle x_k,g_{kl}^tx_l\rangle\geq 0.
$$
\end{enumerate}
\end{theorem}
\begin{proof}
${\it (i)}\Longleftrightarrow {\it (ii)}$ Denote by
$\{k_\alpha\}_{\alpha=1}^k$ and
$\{K_{\alpha\beta}\}_{\alpha,\beta=1}^k$ the standard orthonormal
basis in $\mathbb{C}^k$ and the standard system of matrix units in
$M_k$ respectively. By Theorem 2.1 the map $\varphi_k=\id\otimes
\varphi:M_k\otimes \cB(\bC^m)\to M_k\otimes \cB(\bC^n)$ is positive if and only
if $$ \langle x^{(k)}\otimes y^{(k)},h^{(k)}x^{(k)}\otimes
y^{(k)}\rangle\geq 0 $$ for every $x^{(k)}\in
\mathbb{C}^k\otimes\mathbb{C}^m$ and
$y^{(k)}\in\mathbb{C}^k\otimes \mathbb{C}^n$, where $$
h^{(k)}=\sum_{\alpha,\beta=1}^k\sum_{i,j=1}^m
K_{\alpha\beta}\otimes E_{ij}\otimes
\varphi_k(K_{\alpha\beta}\otimes
E_{ij})=\sum_{\alpha,\beta=1}^k\sum_{i,j=1}^m
K_{\alpha\beta}\otimes E_{ij}\otimes K_{\alpha\beta}\otimes
h_{ij}.$$ Let $x^{(k)}\in \mathbb{C}^k\otimes\mathbb{C}^m$ and
$y^{(k)}\in\mathbb{C}^k\otimes\mathbb{C}^n$, and let
$x_1,\ldots,x_k\in\mathbb{C}^m$, $y_1,\ldots,y_k\in\mathbb{C}^n$
be such that $$ x^{(k)}=\sum_\rho k_\rho\otimes x_\rho,\;\;\;\;
y^{(k)}=\sum_\sigma k_\sigma\otimes y_\sigma. $$ Then
\begin{eqnarray*}
\lefteqn{\langle x^{(k)}\otimes y^{(k)},h^{(k)}x^{(k)}\otimes y^{(k)}\rangle=}\\
&=&\sum_{\rho,\sigma,\rho',\sigma'}\langle k_\rho\otimes x_\rho\otimes k_\sigma\otimes y_\sigma,
  h^{(k)}k_{\rho'}\otimes x_{\rho'}\otimes k_{\sigma'}\otimes y_{\sigma'}\rangle \\
&=&\sum_{\rho,\sigma,\rho',\sigma'}\;\sum_{\alpha,\beta}\,\sum_{i,j}
  \langle k_\rho,K_{\alpha\beta}k_{\rho'}\rangle\langle x_\rho,E_{ij}x_{\rho'}\rangle
  \langle k_\sigma,K_{\alpha\beta}k_{\sigma'}\rangle\langle y_\sigma,h_{ij}y_{\sigma'}\rangle \\
&=&\sum_{\alpha,\beta}\sum_{i,j}\langle x_\beta,E_{ij}x_\alpha\rangle\langle y_\beta,
  h_{ij}y_\alpha\rangle \\
&=&\sum_{\alpha,\beta}\sum_{i,j}\langle e_i,x_\alpha\rangle\langle x_\beta,e_j\rangle\langle
  y_\beta, h_{ij}y_\alpha\rangle \\
&=&\sum_{i,j}\left\langle\sum_\beta\langle e_j,x_\beta\rangle y_\beta,h_{ij}\sum_\alpha
  \langle e_i,x_\alpha\rangle y_\alpha\right\rangle
\end{eqnarray*}
Let $y_i'=\sum_\alpha\langle e_i,x_\alpha\rangle y_\alpha$ for $i=1,\ldots,m$.
Then, the equivalence is obvious.

${\it (ii)}\Longleftrightarrow {\it (iii)}$ This is a consequence
of the following equality:
\begin{eqnarray*}
\lefteqn{\sum_{i,j}\langle y_j,h_{ij}y_i\rangle =
 \sum_{i,j}\sum_{s,t}\langle e_s,E_{ij}e_t\rangle\langle y_s,h_{ij}y_t\rangle} \\
&=&\sum_{s,t}\left\langle e_s\otimes y_s,\left(\sum_{i,j}E_{ij}\otimes h_{ij}\right) e_t
  \otimes y_t\right\rangle
=\sum_{s,t}\langle e_s\otimes y_s,he_t\otimes y_t\rangle \\
&=& \sum_{s,t}\langle e_s\otimes y_s,g^te_t\otimes y_t\rangle 
=\sum_{s,t}\left\langle e_s\otimes y_s,\left(\sum_{k,l}g_{kl}^t\otimes F_{lk}\right) e_t
  \otimes y_t\right\rangle \\
&=& \sum_{s,t}\sum_{k,l}\langle e_s,g_{kl}^te_t\rangle\langle y_s, F_{lk}y_t\rangle 
= \sum_{k,l}\sum_{s,t}\sum_{p,r}\overline{\langle f_p,y_s\rangle}\langle f_r,y_t\rangle
  \langle e_s,g_{kl}^te_t\rangle\langle f_p,F_{lk}f_r\rangle \\
&=& \sum_{k,l}\sum_{s,t}\overline{\langle f_k,y_s\rangle}\langle f_l,y_t\rangle
  \langle e_s,g_{kl}^te_t\rangle 
=\sum_{k,l}\left\langle \sum_s\langle f_k,y_s\rangle e_s, g_{kl}^t\sum_t\langle f_l,y_t
  \rangle e_t\right\rangle
\end{eqnarray*}
Now, define $x_k=\sum_s\langle f_k,y_s\rangle e_s$ for $k=1,\ldots,n$. 
The equivalence follows from the fact that
$$\dim{\rm span}\{x_1,\ldots,x_n\}=\dim{\rm span}\{y_1,\ldots,y_m\}.$$
\end{proof}
As a corollary we get
\begin{theorem}
Let $\varphi:\cB(\bC^m)\to \cB(\bC^n)$ be a linear map. Then the following conditions are
equivalent:
\begin{enumerate}
\item[(i)]
$\varphi$ is $k$-copositive;
\item[(ii)]
for every $y_1,\ldots,y_m\in \mathbb{C}^n$ such that $\dim{\rm span}\{y_1,\ldots,y_m\}\leq k$
we have
$$
\sum_{i,j=1}^n\langle y_i,h_{ij}y_j\rangle\geq 0;
$$
\item[(iii)]
for every $x_1,\ldots,x_n\in\mathbb{C}^m$ such that $\dim{\rm span}\{x_1,\ldots,x_n\}\leq k$
we have
$$
\sum_{k,l=1}^n\langle x_k,g_{kl}x_l\rangle\geq 0.
$$
\end{enumerate}
\end{theorem}
\begin{proof}
With $t$ denoting the transposition map $a \to a^t$, we
let $h'$ and $g'$ denote the operators corresponding to the map
$\varphi\circ t$ in the construction described in Theorems 2.1 and
2.2 for $\varphi$. Then, it is easy to show that $h_{ij}'=h_{ji}$ for every
$i,j=1,\ldots, m$ and $g_{kl}'=g_{kl}^t$ for $k,l=1,\ldots,n$.
Thus, the theorem follows.
\end{proof}

Now, we can generalise this result to the general case. If $H$ is
a Hilbert space let ${\rm Proj}_k(H)=\{p\in \cB(H):\,p^*=p=p^2,\,\Tr
p\leq k\}$. Then we have
\begin{theorem}\label{kth3}
Let $A$ be a $C^*$-algebra, $H$ a Hilbert space (not necessarily
finite dimensional) and $\varphi:A\to \cB(H)$ a linear map. Then the
following conditions are equivalent:
\begin{itemize}
\item[(i)]
$\varphi$ is $k$-positive;
\item[(ii)]
for every $n\in\mathbb{N}$, every set of vectors
$\xi_1,\xi_2,\ldots,\xi_n\in H$ such that $$
\dim\mathrm{span}\{\xi_1,\xi_2,\ldots,\xi_n\}\leq k, $$ and every
$[a_{ij}]\in M_n(A)^+$, we have
$$\sum_{i,j=1}^n\langle\xi_i,\varphi(a_{ij})\xi_j\rangle\geq 0;$$
\item[(iii)]
for every $p\in {\rm Proj}_k(H)$ the map $A\ni a\mapsto p\varphi(a)p\in \cB(H)$
is completely positive.
\end{itemize}
\end{theorem}
\begin{proof}
(i) $\Rightarrow $ (iii) Observe that the map $p\varphi p$ is
$k$-positive as it is a composition of $k$-positive and completely
positive maps. It maps $A$ into $p\cB(H)p$, but the latter
subalgebra is isomorphic with $M_d$ where $d=\Tr p\leq k$. By the
theorem of Tomiyama (\cite{Tom}) $k$-decomposability of $p\varphi
p$ implies its complete positivity.

(iii) $\Rightarrow $ (ii) Let $\xi_1,\xi_2,\ldots,\xi_n\in H$ and
$\dim\mathrm{span}\{\xi_1,\xi_2,\ldots,\xi_n\}\leq k$. If $p$ is a
projection such that
$pH=\mathrm{span}\{\xi_1,\xi_2,\ldots,\xi_n\}$, then $p\in{\rm
Proj}_k(H)$ and hence $p\varphi p$ is completely positive by
assumption. So, for every $[a_{ij}]\in M_n(A)^+$ we have
$$\sum_{i,j=1}^n\langle\xi_i,\varphi(a_{ij})\xi_j\rangle=
\sum_{i,j=1}^n\langle p\xi_i,\varphi(a_{ij})p\xi_j\rangle=
\sum_{i,j=1}^n\langle\xi_i,p\varphi(a_{ij})p\xi_j\rangle \geq 0$$

(ii) $\Rightarrow $ (i) Let $[a_{ij}]\in M_k(A)^+$. Then for every
$\xi_1,\xi_k,\ldots,\xi_k\in H$ we have
$$\sum_{i,j=1}^k\langle\xi_i,\varphi(a_{ij})\xi_j\rangle\geq 0$$
This condition is equivalent to the positivity of the matrix
$[\varphi(a_{ij})]$ in $M_k(\cB(H))$, which implies that $\varphi$ is
$k$-positive.\end{proof}
\begin{corollary}
A map $\varphi:A\to \cB(H)$ is completely positive if and only if
$p\varphi p$ is completely positive for every finite dimensional projector in $\cB(H)$.
\end{corollary}

Now we are ready to study the notion of $k$-decomposability.
\begin{definition}
Let $\varphi:A\to \cB(H)$ be a linear map.
\begin{enumerate}
\item
We say that $\varphi$ is {\em $k$-decomposable}
if there are maps $\varphi_1,\varphi_2:A\to \cB(H)$ such that $\varphi_1$
is $k$-positive, $\varphi_2$ is $k$-copositive and $\varphi=\varphi_1+\varphi_2$.
\item
We say that $\varphi$ is {\em weakly $k$-decomposable} if there is a $C^*$-algebra $E$,
a unital Jordan morphism $\cJ:A\to E$, and a positive map
$\psi :E\to \cB(H)$ such that $\psi|_{\cJ(A)}$ is $k$-positive and
$\varphi=\psi\circ \cJ$.
\end{enumerate}
\end{definition}
\begin{theorem}
For any linear map $\varphi:A\to \cB(H)$ consider the following conditions:
\begin{itemize}
\item[{\rm (D$_k$)}]
$\varphi$ is $k$-decomposable;
\item[{\rm (W$_k$)}]
$\varphi$ is weakly $k$-decomposable;
\item[{\rm (S$_k$)}]
for every matrix $[a_{ij}]\in M_k(A)$ such that both $[a_{ij}]$ and $[a_{ji}]$ are in
$M_k(A)^+$ the matrix $[\varphi(a_{ij})]$ is positive in $M_k(\cB(H))$;
\item[{\rm (P$_k$)}]
for every $p\in {\rm Proj}_k(H)$ the map $p\varphi p$ is decomposable.
\end{itemize}
Then we have the following implications:
{\rm (D$_k$)} $\Rightarrow$ {\rm (W$_k$)} $\Leftrightarrow$ {\rm (P$_k$)}
$\Leftrightarrow$ {\rm (S$_k$)}.
\end{theorem}
\begin{proof}
(D$_k$) $\Rightarrow$ (P$_k$) If $\varphi=\varphi_1+\varphi_2$
with $\varphi_1$ is $k$-positive and $\varphi_2$ $k$-copositive,
then $p\varphi p=p\varphi_1 p+p\varphi_2 p$. From Theorem
\ref{kth3} $p\varphi_1 p$ is a completely positive map. Observe
that $p^t\in{\rm Proj}_k(H)$ for every $p\in{\rm Proj}_k(H)$.
Hence $(p\varphi_2 p)^t=p^t\varphi_2^tp^t$ and $(p\varphi_2p)^t$
is completely positive. Thus $p\varphi p$ is a sum of a completely
positive and completely copositive map, and hence $p\varphi p$ is
decomposable.

(P$_k$) $\Rightarrow$ (S$_k$) Let $[a_{ij}]\in M_k(A)$ be such
that $[a_{ij}],[a_{ji}]\in M_k(A)^+$. Suppose that
$\xi_1,\xi_2,\ldots,\xi_k\in H$ and that $p$ is a projector on $H$
such that $pH=\mathrm{span}\{\xi_1,\xi_2,\ldots,\xi_k\}$. Then
$$\sum_{i,j=1}^k\langle\xi_i,\varphi(a_{ij})\xi_j\rangle=
\sum_{i,j=1}^k\langle p\xi_i,\varphi(a_{ij})p\xi_j\rangle=
\sum_{i,j=1}^k\langle\xi_i,p\varphi(a_{ij})p\xi_j\rangle\geq 0$$
where in the last inequality we have used the fact that the matrix
$[p\varphi(a_{ij})p]$ is positive by the theorem of St{\o}rmer.
Hence the matrix $[\varphi(a_{ij})]$ is positive.

(S$_k$) $\Rightarrow$ (P$_k$)
Let $p\in{\rm Proj}_k(H)$ and $d=\Tr p$. One should show that for every
$n\in\mathbb{N}$ and every matrix $[a_{ij}]\in M_n(A)$ such that $[a_{ij}],[a_{ji}]
\in M_n(A)^+$ the matrix $[p\varphi(a_{ij})p]$ is also positive. To this end
we will show that for any vectors $\xi_1,\xi_2,\ldots,\xi_n$
the inequality
\begin{equation}\label{k1}
\sum_{i,j=1}^n\langle\xi_i,p\varphi(a_{ij})p\xi_j\rangle\geq 0
\end{equation}
holds.
If $n\leq k$ then we define vectors $\eta_1,\eta_2,\ldots,\eta_k$:
$$\eta_i=\left\{\begin{array}{ll}
p\xi_i&\mbox{for $1\leq i\leq n$,}\\
0&\mbox{for $n<i\leq k$}\end{array}\right.$$
and a matrix $[b_{ij}]\in M_k(A)$:
$$b_{ij}=\left\{\begin{array}{ll}
a_{ij}&\mbox{for $1\leq i,j\leq n$,}\\
0&\mbox{otherwise.}
\end{array}\right.$$
Obviously both matrices $[b_{ij}]$ and $[b_{ji}]$ are positive in $M_k(A)$.
Thus
$$\sum_{i,j=1}^n\langle\xi_i,p\varphi(a_{ij})p\xi_j\rangle=
\sum_{i,j=1}^k\langle\eta_i,\varphi(b_{ij})\eta_j\rangle\geq 0$$
by assumption.
Now, let us assume that $n=k+1$. Define $\eta_i=p\xi_i$ for $i=1,2,\ldots,k+1$.
As $\dim\mathrm{span}\{\eta_1,\eta_2,\ldots,\eta_{k+1}\}\leq k$ then at least one of
vectors $\eta_1,\eta_2,\ldots,\eta_{k+1}$, say $\eta_{k+1}$, is a linear combination
of the others, i.e. $\eta_{k+1}=\sum_{i=1}^k\alpha_i\eta_i$. Then
\begin{eqnarray*}
\lefteqn{\sum_{i,j=1}^{k+1}\langle\xi_i,p\varphi(a_{ij})p\xi_j\rangle=
\sum_{i,j=1}^{k+1}\langle\eta_i,\varphi(a_{ij})\eta_j\rangle=}\\
&=&
\sum_{i,j=1}^{k}\langle\eta_i,\varphi(a_{ij})\eta_j\rangle+
\sum_{i=1}^{k}\langle\eta_i,\varphi(a_{i,k+1})\eta_{k+1}\rangle +\\
&&+
\sum_{j=1}^{k}\langle\eta_{k+1},\varphi(a_{k+1,j})\eta_j\rangle+
\langle\eta_{k+1},\varphi(a_{k+1,k+1})\eta_{k+1}\rangle =\\
&=&
\sum_{i,j=1}^{k}\langle\eta_i,\varphi(a_{ij})\eta_j\rangle+
\sum_{i,j=1}^{k}\langle\eta_i,\alpha_j\varphi(a_{i,k+1})\eta_j\rangle +\\
&&+
\sum_{i,j=1}^{k}\langle\alpha_i\eta_i,\varphi(a_{k+1,j})\eta_j\rangle+
\sum_{i,j=1}^{k}\langle\alpha_i\eta_i,\alpha_j\varphi(a_{k+1,k+1})\eta_j\rangle =\\
&=&
\sum_{i,j=1}^{k}\langle\eta_i,
\left[\varphi(a_{ij})+
\alpha_j\varphi(a_{i,k+1})+
\overline{\alpha_i}\varphi(a_{k+1,j})+
\overline{\alpha_i}\alpha_j\varphi(a_{k+1,k+1})\right]\eta_j\rangle =\\
&=&\sum_{i,j=1}^k\langle\eta_i,\varphi(b_{ij})\eta_j\rangle
\end{eqnarray*}
where $b_{ij}=a_{ij}+\alpha_ja_{i,k+1}+\overline{\alpha_i}a_{k+1,j}+
\overline{\alpha_i}\alpha_ja_{k+1,k+1}$ for $i,j=1,2,\ldots,k$.
The fact that both
matrices $[b_{ij}]$ and $[b_{ji}]$ are positive in $M_k(A)$, follows
from the following matrix equality
\begin{eqnarray*}
\lefteqn{\left[\begin{array}{ccccc}
b_{11}&b_{12}&\cdots&b_{1k}&0\\
b_{21}&b_{22}&\cdots&b_{2k}&0\\
\cdot&\cdot& &\cdot&\cdot\\
\cdot&\cdot& &\cdot&\cdot\\
b_{k1}&b_{k2}&\cdots&b_{kk}&0\\
0&0&\cdots&0&0
\end{array}\right] }\\
&=&
\left[\begin{array}{ccccc}
1&0&\cdots&0&\overline{\alpha_1}\\
0&1&\cdots&0&\overline{\alpha_2}\\
\cdot&\cdot& &\cdot&\cdot\\
\cdot&\cdot& &\cdot&\cdot\\
0&0&\cdots&1&\overline{\alpha_k}\\
0&0&\cdots&0&0
\end{array}\right]
[a_{ij}]
\left[\begin{array}{ccccc}
1&0&\cdots&0&0\\
0&1&\cdots&0&0\\
\cdot&\cdot& &\cdot&\cdot\\
\cdot&\cdot& &\cdot&\cdot\\
0&0&\cdots&1&0\\
\alpha_1&\alpha_2&\cdots&\alpha_k&0
\end{array}\right]
\end{eqnarray*}
Hence, by assumption inequality (\ref{k1}) holds.
We may continue the proof for larger $n$ by a similar inductive argument.

(W$_k$) $\Leftrightarrow $ (S$_k$) We follow the proof of the
Theorem in \cite{Sto2}. For the reader's convenience we describe
St{\o}rmer's argument:

(W$_k$) $\Rightarrow $ (S$_k$) If $\cJ$ is a $*$-homomorphism
(resp. $*$-antihomomorphism) and $[a_{ij}]$ (resp. $[a_{ji}]$) is
in $M_k(A)^+$ then $[\cJ(a_{ij})]$ belongs to $M_k(E)^+$. Since
every Jordan morphism is a sum of a $*$-homomorphism and a
$*$-antimorphism, if both $[a_{ij}]$ and $[a_{ji}]$ belong to
$M_k(A)^+$ then $[\cJ(a_{ij})]\in M_k(\cB(H))^+$. Applying $\psi$
now yields the fact that $[\varphi(a_{ij})]\in M_k(\cB(H))^+$.

(S$_k$) $\Rightarrow $ (W$_k$) Assume that $A\subset \cB(L)$ for
some Hilbert space $L$. Let $$
V=\left\{\left[\begin{array}{cc}a&0\\0&a^{t'}\end{array}\right]\in
M_2(\cB(L)):\, a\in A\right\} $$ where $t'$ is the transposition map
with respect to some orthonormal basis in $L$. Then $V$ is a
selfadjoint subspace of $M_2(\cB(L))$ containing the identity. One
can observe that both $[a_{ij}]$ and $[a_{ji}]$ belong to
$M_k(A)^+$ if and only if $$ \left[\begin{array}{ccc}
\left[\begin{array}{cc}a_{11}&0\\0&a_{11}^{t'}\end{array}\right]&\ldots&
\left[\begin{array}{cc}a_{1k}&0\\0&a_{1k}^{t'}\end{array}\right]\\
\cdot&&\cdot\\ \cdot&&\cdot\\
\left[\begin{array}{cc}a_{k1}&0\\0&a_{k1}^{t'}\end{array}\right]&\ldots&
\left[\begin{array}{cc}a_{kk}&0\\0&a_{kk}^{t'}\end{array}\right]
\end{array}\right]\in M_k(V)^+.
$$
Thus the map $\psi:V\to \cB(H)$ defined by
\begin{equation}\label{k2}
\psi\left(\left[\begin{array}{cc}a&0\\0&a^{t'}\end{array}\right]\right)=
\varphi(a)
\end{equation}
is $k$-positive. Now, take $E=M_2(\cB(L))$ and define the Jordan
morphism $\cJ:A\to M_2(\cB(L))$ by $$
\cJ(a)=\left[\begin{array}{cc}a&0\\0&a^{t'}\end{array}\right] $$
to prove the statement.
\end{proof}

We end this section with the remark that it is still an open
problem if conditions (S$_k$), (P$_k$) and (W$_k$) are equivalent
to $k$-decomposability. The main difficulty in proving the
implication, say (S$_k$) $\Rightarrow $ (D$_k$), is to find a
$k$-positive extension of the map $\psi$ constructed in (\ref{k2})
to the whole algebra $M_2(\cB(L))$. So, one should answer the
following question:
\begin{quote}
Given a $C^*$-algebra $A$ and a selfadjoint linear unital subspace
$S$, find conditions for $k$-positive maps $\psi:S\to \cB(H)$ which
guarantee the existence of a $k$-positive extension of $\psi$ to
whole algebra $A$.
\end{quote}
In other words, the analog of Arveson's extension theorem for 
completely positive maps should be proved (\cite{Arv}, see also 
\cite{Sto4}). The results concerning this problem will be included 
in the forthcoming paper \cite{LMM}.

\section{Tomita-Takesaki scheme for transposition}\label{Tomita}
Let $H$ be a finite dimensional (say $n$-dimensional) Hilbert space. 
We are concerned with a strongly positive map $\varphi:\cB(H)\to \cB(H)$, 
i.e. a map such that $\varphi(a^*a) \geq \varphi(a)^*\varphi(a)$ for 
every $a\in \cB(H)$ (also called a Schwarz map).

Define $\omega\in \cB(H)_{+,1}^*$ as
$
\omega(a)= \Tr\varrho a,
$
where $\varrho$ is an invertible density matrix, i.e. the state $\omega$
is a faithful one.
Denote by $(H_\pi,\pi,\Omega)$ the GNS triple associated with
$(\cB(H),\omega)$.
Then, one has:
\begin{itemize}
\item
$H_\pi$ is identified with $\cB(H)$ where the inner product
$(\cdot\,,\cdot)$ is defined
as $(a,b)=\Tr a^*b$, $a,b\in \cB(H)$;
\item
With the above identification: $\Omega= \rr$;
\item
$\pi(a)\Omega=a\Omega$;
\item
The modular conjugation $\Jm$ is the hermitian involution:
$\Jm a \rr =\rr a^*$;
\item
The modular operator $\Delta$ is equal to the map
$\varrho \cdot\varrho^{-1}$;
\end{itemize}
We assume that $\omega$ is invariant with respect to $\varphi$,
i.e. $\omega\circ\varphi=\omega$. Now, let us consider the
operator $T_{\varphi}\in \cB(H_\pi)$ defined by 
$$
T_{\varphi}
(a\Omega)=\varphi(a)\Omega,\;\;\;a\in \cB(H). 
$$ 
Obviously
$T_{\varphi}$ is a contraction due to the strong positivity of
$\varphi$. 

As a next step let us define two conjugations: $J_c$ on
$H$ and $J$ on $H_\pi$.
To this end we note that the eigenvectors $\{x_i\}$ of $\varrho
=\sum_i \lambda_i |x_i\rangle \langle x_i|$ form an orthonormal basis in $H$
(due to the faithfulness of $\omega$). Hence we can define
\begin{equation}\label{Jcdef}
J_c f = \sum_i \overline{\langle x_i,f\rangle} x_i
\end{equation}
for every $f\in H$.
Due to the fact that $E_{ij}\equiv |x_i\rangle \langle x_j|\}$ form an 
orthonormal basis in $H_\pi$ we can define in the similar way a conjugation $J$ on
$H_\pi$
\begin{equation}\label{Jdef}
J a \rr = \sum_{ij} \overline{(E_{ij},a \rr)} E_{ij}
\end{equation}
Obviously, $J\rr = \rr$. 

Now let us define a transposition on $\cB(H)$ as the map $a\mapsto a^t\equiv J_ca^*J_c$ where
$a\in \cB(H)$. By $\tau$ we will denote the map induced on $H_\pi$ by the transposition, i.e.
\begin{equation}\label{tau}
\tau a\rr=a^t\rr
\end{equation}
where $a\in \cB(H)$.
The main properties of the notions introduced above are the following
\begin{proposition}\label{transposition}
Let $a\in \cB(H)$ and $\xi\in H_\pi$. Then
$$a^t\xi=Ja^*J\xi.$$
\end{proposition}
\begin{proof}
Let $\xi=b\rr$ for some $b\in \cB(H)$. Then we can perform the following cal\-cu\-la\-tions
\begin{eqnarray*}
\lefteqn{Ja^*Jb\rr=}\\
&=&
\sum_{ij}\overline{(E_{ij},a^*Jb\rr)}E_{ij}=
\sum_{ij}\sum_{kl}(E_{kl},b\rr)\overline{(E_{ij},a^*E_{kl})}E_{ij}\\
&=&
\sum_{ijkl}\Tr (E_{lk}b\rr)\overline{\Tr (E_{ji}a^*E_{kl})}E_{ij}=
\sum_{ijk}\Tr (E_{jk}b\rr)\overline{\Tr (E_{ki}a^*)}E_{ij}\\
&=&
\sum_{ijk}\langle x_k,b\rr x_j\rangle\overline{\langle x_i,a^*x_k\rangle}E_{ij}=
\sum_{ijk}\langle J_cb\rr x_j,x_k\rangle\langle x_k,ax_i\rangle E_{ij}\\
&=&
\sum_{ij}\langle J_cb\rr x_j,ax_i\rangle E_{ij}=
\sum_{ij}\langle a^*J_cb\rr x_j,x_i\rangle E_{ij}\\
&=&
\sum_{ij}\langle x_i,J_ca^*J_cb\rr x_j\rangle E_{ij}=
\sum_{ij}\langle x_i,a^tb\rr x_j\rangle E_{ij}\\
&=&
\sum_{ij}\Tr (E_{ji}a^tb\rr)E_{ij}=
\sum_{ij}(E_{ij},a^tb\rr)E_{ij}=
a^tb\rr
\end{eqnarray*}
\end{proof}

As a next step let us consider the modular conjugation $\Jm$ which
has the form
\begin{equation}\label{Jm}
\Jm a \rr = (a\rr)^* = \rr a^*
\end{equation}
Define also the unitary operator $U$ on $H_\pi$ by
\begin{equation}\label{U}
U = \sum_{ij} |E_{ji}\rangle \langle E_{ij}|
\end{equation}
Clearly, $UE_{ij} = E_{ji}$. We have the following
\begin{proposition}\label{commute}
Let $J$ and $\Jm$ be the conjugations introduced above and $U$ be the unitary operator
defined by (\ref{U}). Then we have:
\begin{enumerate}
\item\label{commute1} $U^2 = \jed$ and $U = U^*$
\item\label{commute2} $J=U\Jm$;
\item\label{commute3} $J$, $\Jm$ and $U$ mutually commute.
\end{enumerate}
\end{proposition}
\begin{proof}
(\ref{commute1})
We calculate
$$ \sum_{ijmn}|E_{ij}\rangle \langle E_{ji}||E_{mn}\rangle \langle E_{nm}|
=\sum_{ijmn} \Tr (E_{ij} E_{mn}) |E_{ij}\rangle \langle E_{nm}|
=\sum_{ij} |E_{ij}\rangle \langle E_{ij}| = \jed
$$
The rest is evident.

(\ref{commute2}) 
Let $b\in \cB(H)$. Then 
\begin{eqnarray*}
U\Jm b\rr&=&U\rr b^*=\sum_{ij}(E_{ji},\rr b^*)E_{ij}\\
&=&\sum_{ij}\Tr (E_{ij}\rr b^*)E_{ij}=
\sum_{ij}\langle x_j,\rr b^*x_i\rangle E_{ij}\\
&=&\sum_{ij}\overline{\langle x_i,b\rr x_j\rangle}E_{ij}=
\sum_{ij}\overline{\Tr (E_{ji}b\rr )}E_{ij}\\
&=&\sum_{ij}\overline{(E_{ij},b\rr)}E_{ij}=
Jb\rr
\end{eqnarray*}

(\ref{commute3}) 
$J$ is an involution, so by the previous point we have $U\Jm U\Jm=\jed$.
It is equivalent to the equality $U\Jm=\Jm U$. Hence we obtain $U\Jm=J=\Jm U$ and
consequently 
$UJ=\Jm=JU$ and $\Jm J=U=J\Jm$ because both $U$ and $\Jm$ are also 
involutions.

\end{proof}

Now, we are ready to describe a polar decomposition of the map $\tau$.
\begin{theorem}\label{polar}
If $\tau$ is the map introduced in (\ref{tau}), then
$$
\tau=U\Delta^{1/2}.
$$
\end{theorem}
\begin{proof}
Let $a\in \cB(H)$. Then by Proposition \ref{transposition} and Proposition \ref{commute}(2)
we have
$$
\tau a\rr = a^t \rr = J a^* J \rr = J \Jm  \Delta^{1/2}a \rr
= U \Delta^{1/2} a \rr.
$$
\end{proof}

Now we wish to prove some properties of $U$ which are analogous to that of the
modular conjugation $\Jm$. To this end we firstly need the following
\begin{lemma}\label{JcomD}
$J$ commutes with $\Delta$
\end{lemma}
\begin{proof}
Let $a\in \cB(H)$. Then by Propositions \ref{transposition}, \ref{commute} 
and Theorem \ref{polar} we have
\begin{eqnarray*}
\Delta^{1/2}Ja\rr&=&\Delta^{1/2}JaJ\rr=
\Delta^{1/2}(a^*)^t\rr=UU\Delta^{1/2}(a^*)^t\rr\\
&=&Ua^*\rr=UJJa^*J\rr=
JUa^t\rr\\&=&JUU\Delta^{1/2}a\rr=
J\Delta^{1/2}a\rr
\end{eqnarray*}
So, $\Delta^{1/2}J=J\Delta^{1/2}$ and consequently $\Delta J=\Delta^{1/2}J\Delta^{1/2}=
J\Delta$.
\end{proof}
We will also use (cf. \cite{Ara}) 
$$ 
V_{\beta} = \mathrm{closure} \left\{
\Delta^{\beta}a \rr:\; a \geq 0, \;\beta \in
\left[0,\frac{1}{2}\right]\right\}. 
$$ 
Clearly, each $V_{\beta}$
is a pointed, generating cone in $H_\pi$ and 
\begin{equation}\label{duality}
V_\beta=\{\xi\in H_\pi:\,\mbox{$(\eta,\xi)\geq 0$ for all $\eta\in V_{(1/2)-\beta}$}\}
\end{equation}
Recall that $V_{1/4}$ is nothing
but the natural cone $\cP$ associated with the pair $(\pi(\cB(H)),\Omega)$
(see \cite[Proposition 2.5.26(1)]{BR}).
Finally, let us define an automorphism $\alpha$ on $\cB(H_\pi)$ by
\begin{equation}\label{alfa}
\alpha(a) = U a U^*,\quad a\in \cB(H_\pi).
\end{equation}
Then we have
\begin{proposition}
\label{przestawianie}
\begin{enumerate}
\item\label{przestawianie1} $U \Delta = \Delta^{-1} U$
\item\label{przestawianie2} $\alpha$ maps $\pi(\cB(H))$ onto $\pi(\cB(H))^{\prime}$;
\item\label{przestawianie3} For every $\beta\in [0,1/2]$ the unitary $U$ maps $V_\beta$ 
onto $V_{(1/2)-\beta}$.
\end{enumerate}
\end{proposition}
\begin{proof}
(\ref{przestawianie1})
By Proposition \ref{commute} and Lemma \ref{JcomD} we have
$$
U\Delta=J\Jm\Delta=J\Delta^{-1}\Jm=\Delta^{-1}J\Jm.
$$

(\ref{przestawianie2})
Let $a,b\in \cB(H)$ and $\xi\in H_\pi$. Then Propositions \ref{transposition} and 
\ref{commute} imply
\begin{eqnarray*}
UaUb\xi&=&J\Jm a\Jm JbJJ\xi=J\Jm a\Jm (b^*)^tJ\xi=J(b^*)^t\Jm a\Jm J\xi\\
&=&J(b^*)^tJJ\Jm a\Jm J\xi=bJ\Jm a\Jm J\xi=bUaU\xi
\end{eqnarray*}
and the proof is complete.

(\ref{przestawianie3})
Let $a,b\in \cB(H)^+$. Then by the point (\ref{przestawianie1}) and Theorem \ref{polar} 
we have
\begin{eqnarray*}
\lefteqn{(\Delta^\beta b \rr, U \Delta^\beta a \rr)=}\\
&=& (\Delta^\beta b\rr, \Delta^{(1/2)-\beta}U\Delta^{1/2} a\rr)=
(\Delta^\beta b\rr, \Delta^{(1/2)-\beta} a^t\rr)
\end{eqnarray*}
We recall that $a\mapsto a^t$ is a positive map on $\cB(H)$ so
by (\ref{duality}) the last expression is nonnegative.
Hence $UV_\beta\subset V_{(1/2)-\beta}$ for every $\beta\in [0,1/2]$.
As $U$ is an involution, we get $V_{(1/2)-\beta}=U^2V_{(1/2)-\beta}\subset
UV_\beta$ and the proof is complete.
\end{proof}

\begin{corollary}
$U \Delta^{1/2}$ and $T_{\varphi} U \Delta^{1/2}$
map $V_0$ into itself.
\end{corollary}

Summarizing, this section establishes a close relationship between
the Tomita-Takesaki scheme and transposition.
Moreover, we have the following :
\begin{proposition}
Let $\xi\mapsto\omega_\xi$ be the homeomorphism between the natural
cone $\cP$ and the set of normal states on $\pi(\cB(H))$ described in
\cite[Theorem 2.5.31]{BR}, i.e. such that
$$\omega_\xi(a)=(\xi,a\xi),\quad a\in \cB(H).$$
For every state $\omega$ define
$\omega^\tau(a)=\omega(a^t)$ where $a\in \cB(H)$.
If $\xi\in\cP$ then the unique vector in $\cP$ mapped 
into the state
$\omega_\xi^\tau$ by the homeomorphism described above, 
is equal to $U\xi$
\end{proposition}
\begin{proof}
Let $\xi=\Delta^{1/4}a\Omega$ for some $a\in \cB(H)^+$. Then we have
\begin{eqnarray*}
(U \Delta^{\frac{1}{4}} a \Omega, x U \Delta^{\frac{1}{4}}a \Omega)
&=& (\Delta^{\frac{1}{4}} U \Delta^{\frac{1}{2}}a \Omega, x \Delta^{\frac{1}{4}}
U \Delta^{\frac{1}{2}} a \Omega) \\
&=& (\Delta^{\frac{1}{4}} a^t \Omega, x \Delta^{\frac{1}{4}} a^t \Omega)\\
&=&(\Delta^{\frac{1}{4}} JaJ \Omega, x \Delta^{\frac{1}{4}} J a J \Omega)\\
&=& (x^*J \Delta^{\frac{1}{4}} a \Omega, J \Delta^{\frac{1}{4}} a \Omega)\\
&=&(\Delta^{\frac{1}{4}}a \Omega, Jx^*J \Delta^{\frac{1}{4}} a \Omega)
\end{eqnarray*}
\end{proof}

\section{$k$-decomposability at the Hilbert-space level}\label{hilbert}
The results of Section \ref{Tomita} strongly suggest that a more
complete theory of $k$-decomposable maps may be obtained in
Hilbert-space terms. To examine that question we will study the
description of positivity in the dual approach to that given in
Section \ref{kdec}, i.e. we will be concerned with the approach on
the Hilbert space level.

Let ${\mathcal{M}} \subset \cB(H)$ be a concrete von Neumann algebra
with a cyclic and separating vector $\Omega$. When used, $\omega$
will denote the vector state $\omega = (\Omega, \cdot\Omega)$. The
natural cone (modular operator) associated with $({\mathcal M},
\Omega)$ will be denoted by $\mathcal P$ ($\Delta$ respectively).

By $\mathcal{P}_n$ we denote the natural cone for $(\mathcal{M}
\otimes \cB(\mathbb{C}^n), \omega \otimes \omega_0 )$ where $\omega_0$
is a faithful state on $ \cB(\mathbb{C}^n)$ (as an example of $\omega_0$ one can take 
$\frac{1}{n} Tr$). For
the same algebra, $\Delta_n = \Delta \otimes \Delta_0$ and $J_n$
being respectively the modular operator and modular conjugation
for $M_n(\mathcal{M})$, are defined in terms of the vector
$\Omega_n = \Omega \otimes \Omega_0$ (ie. in
terms of the state $\omega \otimes \omega_0$).

We will consider unital positive maps $\varphi$ on $\mathcal{M}$
which satisfy Detailed Balance II, i.e. there is another positive
unital map $\varphi^{\beta}$ such
$\omega(a^*\varphi(b))=\omega(\varphi^{\beta}(a^*)b)$ (see
\cite{MS}). Such maps induce bounded maps $T_{\varphi} = T$ on
$H_{\omega} = H$ which commute strongly with $\Delta$ and which
satisfy $T^*(\mathcal{P}) \subset \mathcal{P}$. Now under the
above assumptions (\cite{wam}; Lemma 4.10) assures us that this
correspondence is actually 1-1. Partial transposition
$({\mathrm{id}}\otimes \tau)$ on $M_n(\mathcal{M})$ also induces
an operator at the Hilbert space level, but for the sake of
simplicity we will where convenient retain the notation
$({\mathrm{id}}\otimes \tau)$ for this operator.

In order to achieve the desired classification of positive maps we
introduce the notion of the ``transposed cone''
$\mathcal{P}_{n}^{\tau} = ({\mathrm{id}} \otimes U)
\mathcal{P}_n$, where $\tau$ is transposition on $M_n(\mathbb{C})$
while the operator $U$ was defined in the previous Section (we
have used the following identification: for the basis $\{e_i \}_i$
in $\mathbb{C}^n$ consisting of eigenvectors of $\varrho_{\omega_0}$
($\omega_0(\cdot) = Tr\{ \varrho_{\omega_0} \cdot \}$, we have the
basis $\{ E_{ij} \equiv |e_i><e_j|\}_{ij}$ in the GNS Hilbert
space associated with $(\cB(\mathbb{C}^n), \omega_0)$ with
$U$ defined in terms of that basis). Note that in the same basis one has 
the identification $\cB(\mathbb{C}^n)$ with  $M_n(\mathbb{C})$.

Now the natural cone $\mathcal{P}_n$ for $\mathcal{M} \otimes
\cB(\mathbb{C}^n) = M_n(\mathcal{M})$ may be realised as
$$\mathcal{P}_n = \overline{{\Delta}^{1/4}_n\{[a_{ij}]\Omega_n :
[a_{ij}] \in M_n(\mathcal{M})^+\}}$$ 
(see for example \cite[Proposition 2.5.26]{BR}). We observe:
\begin{eqnarray*}
\lefteqn{
\{({\mathbb{I}}\otimes U){\Delta}^{1/4}_n [a_{ij}]\Omega_n :\; [a_{ij}] \in
M_n(\mathcal{M})^+\}
}\\
&=& \{ ({\Delta}^{1/4} \otimes U {\Delta}^{1/4}_0) \circ \sum a_{ij}\otimes E_{ij}:\,
    [a_{ij}] \in M_n({\mathcal{M}})^+\}\\
&=& \{({\Delta}^{1/4} \otimes {\Delta}^{1/4}_0 U {\Delta}^{1/2}_0) \circ \sum a_{ij}
    \otimes E_{ij}:\, [a_{ij}] \in M_n({\mathcal{M}})^+\}\\
&=& \{{\Delta}^{1/4}[a_{ji}]\Omega_n :\, [a_{ij}] \in M_n(\mathcal{M})^+\}.
\end{eqnarray*}
Thus
$$\mathcal{P}_n^{\tau}
= \overline{{\Delta}^{1/4}_n\{[a_{ji}]\Omega_n :
[a_{ij}] \in M_n(\mathcal{M})^+\}}.$$
The task of describing the
transposed cone will be addressed more adequately in the next
section.

\begin{lemma}
The map $\varphi : \mathcal{M} \rightarrow \mathcal{M}$ is
$k$-positive ($k$-copositive) if and only if $(T_{\varphi} \otimes
\mathbb{I})^*(\mathcal{P}_n) \subset \mathcal{P}_n$ (respectively
$(T_{\varphi} \otimes \mathbb{I})^*(\mathcal{P}_n) \subset
\mathcal{P}_n^{\tau}$) for every $1 \leq n \leq k$.
\end{lemma}
\begin{proof}
To prove $k$-positivity case it is enough to
suitably adapt the proof of (\cite{wam}; Lemma 4.10),
while to prove $k$-copositivity we observe that the ``if'' part of the hypothesis implies
$$0 \le ( (T_{\varphi} \otimes \mathbb{I})(\mathbb{I} \otimes U)\cP_n, \cP_n).$$
Thus
$$(\Delta_n^{1/4} ([T(a_{ji})]) \Omega_n, \Delta^{-{1/4}}_n ([b_{kl}]^*[b_{kl}]) \Omega_n)
= ([T(a_{ji})]\Omega_n, [b_{kl}]^*[b_{kl}] \Omega_n) \ge 0$$
where $[a_{ij}] \ge 0$ is in the algebra $M_n({\mathcal M})$, and $[b_{kl}]$ in its commutant.
This implies $[T(a_{ji})] \ge 0$ and the rest is again a suitable adaptation of the proof of
(\cite{wam}; Lemma 4.10). 
\end{proof}

\begin{lemma}
For each $n$, $\mathcal{P}_n \cap \mathcal{P}_n^{\tau}$ and
$\overline{co}(\mathcal{P}_n \cup \mathcal{P}_n^{\tau})$ are dual
cones.
\end{lemma}
\begin{proof}
For any $X\subset H$ we denote
$X^{\mathrm{d}}=\{\xi\in H:\,\mbox{$(\xi,\eta)\geq 0$ for any $\eta\in X$}\}$.
To prove the lemma it is enough to observe that
${\mathcal{P}}_n^{\mathrm{d}}={\mathcal{P}}_n$ and
$({\mathcal{P}}_n^{\tau})^{\mathrm{d}}={\mathcal{P}}_n^{\tau}$.
\end{proof}

\begin{lemma}
\label{aa}
Let $n$ be given. For any $[a_{ij}] \in M_n(\mathcal{M})^+$,
$\Delta_n^{1/4}[a_{ij}]\Omega_n \in \mathcal{P}_n \cap
\mathcal{P}_n^{\tau}$ implies $[a_{ji}] \in M_n(\mathcal{M})^+$.
\end{lemma}
\begin{proof}
Let $[a_{ij}] \in M_n(\mathcal{M})^+$ be given and assume that
$\Delta_n^{1/4}[a_{ij}]\Omega_n \in \mathcal{P}_n \cap
\mathcal{P}_n^{\tau}$. We observe
$$\Delta_n^{1/4}[a_{ji}]\Omega_n = ({\mathbb{I}} \otimes
U)\Delta_n^{1/4}[a_{ij}]\Omega_n \in ({\mathbb{I}} \otimes
U)(\mathcal{P}_n \cap \mathcal{P}_n^{\tau}) = \mathcal{P}_n
\cap \mathcal{P}_n^{\tau} \subset \mathcal{P}_n.$$ But then the
self-duality of $\mathcal{P}_n$ alongside (\cite{BR}; 2.5.26) will
ensure that $$0 \leq (\Delta_n^{1/4}[a_{ji}]\Omega_n,
\Delta_n^{-1/4}[b_{ij}]\Omega_n) = ([a_{ji}]\Omega_n,
[b_{ij}]\Omega_n)$$ for each $[b_{ij}] \in (M_n(\mathcal{M})')^+$.
We may now conclude from (\cite{Di}; 2.5.1 or \cite{BR}; 2.3.19)
that $[a_{ji}] \geq 0$, as required.
\end{proof}

\begin{corollary}
\label{wniosek1}
In the finite dimensional case $\{\Delta_n^{1/4}[a_{ij}]\Omega_n :
[a_{ij}] \geq 0, [a_{ji}] \geq 0\} = \mathcal{P}_n \cap
\mathcal{P}_n^{\tau}$.
\end{corollary}
\begin{proof}
First note that in this case $\{\Delta_n^{1/4}[a_{ij}]\Omega_n :
[a_{ij}] \geq 0\} = \mathcal{P}_n$ (cf. \cite[Proposition 2.5.26]{BR}).
Now apply the previous lemma.
\end{proof}

Recall $\Delta_n^{-1/4}$ maps $\overline{\{[b_{ij}]\Omega_n :
[b_{ij}] \in (M_n(\mathcal{M})')^+\}}$ densely into
$\mathcal{P}_n$ (see for example \cite{BR}). At least on a formal
level one may therefore by analogy with 
\cite[2.5.26 \&
2.5.27]{BR}
expect to end up with a dense subset of
the dual cone of $\overline{co}(\mathcal{P}_n \cup
\mathcal{P}_n^\tau)$ (ie. of $\mathcal{P}_n \cap
\mathcal{P}_n^\tau$) when applying $\Delta_n^{1/4}$ to the set of
all $\alpha$'s satisfying
$([b_{ij}]\Omega_n, \alpha) \geq 0$ and $([b_{ij}]\Omega_n,
(\mathbb{I} \otimes U)\alpha) \geq 0$ for each $[b_{ij}] \in
(M_n(\mathcal{M})')^+$. If true such a fact would then put one in
a position to try and show that in general $\mathcal{P}_n \cap
\mathcal{P}_n^{\tau} = \overline{\{\Delta_n^{1/4}[a_{ij}]\Omega_n
: [a_{ij}] \geq 0, [a_{ji}] \geq 0\}}$.

\textbf{Question.} Is it generally true that
$$\overline{\{\Delta_n^{1/4}[a_{ij}]\Omega_n : [a_{ij}] \geq 0,
[a_{ji}] \geq 0\}} = \mathcal{P}_n \cap \mathcal{P}_n^{\tau}?$$
In the light of the following result the answer to this becomes
important in an attempt to generalize the finite case to the infinite
dimensional one.

\begin{theorem}\label{dual}
In general the property $(T_{\varphi} \otimes
\mathbb{I})^*(\mathcal{P}_n) \subset \overline{co}(\mathcal{P}_n
\cup \mathcal{P}_n^{\tau})$ for each $1 \leq n \leq k$ implies
that $\varphi$ is weakly $k$-decomposable in the sense that for
each $1 \leq n \leq k$, $[\varphi(a_{ij})] \geq 0$ whenever
$[a_{ij}], [a_{ji}] \in M_n(\mathcal{M})^+$.

If $\overline{\{\Delta_n^{1/4}[a_{ij}]\Omega_n : [a_{ij}] \geq 0,
[a_{ji}] \geq 0\}} = \mathcal{P}_n \cap \mathcal{P}_n^{\tau}$ for
each $1 \leq n \leq k$, the converse implication also holds. In
particular in the finite-dimensional case the two statements are
equivalent. (Pending the answer to the aforementioned question,
they may of course be equivalent in general.)
\end{theorem}
\begin{proof}
Suppose that $(T_{\varphi} \otimes
\mathbb{I})^*(\mathcal{P}_n) \subset \overline{co}(\mathcal{P}_n
\cup \mathcal{P}_n^{\tau})$ for each $1 \leq n \leq k$. Given
$1\leq n \leq k$ and $[a_{ij}] \in M_n(\mathcal{M})$ it now follows
from \cite[Proposition 2.3.19]{BR} and the strong commutation of
$T_{\varphi} \otimes \mathbb{I}$ with $\Delta_k$, that
$[\varphi(a_{ij})] \geq 0$ if and only if
\begin{eqnarray*}
0 &\leq& (\varphi_n([a_{ij}])\Omega_n, [b_{ij}]\Omega_n)\\ &=&
([a_{ij}]\Omega_n, (T_{\varphi} \otimes
\mathbb{I})^*[b_{ij}]\Omega_n)\\ &=&
(\Delta_n^{1/4}[a_{ij}]\Omega_n, (T_{\varphi} \otimes
\mathbb{I})^*\Delta_n^{-1/4}[b_{ij}]\Omega_n)
\end{eqnarray*}
for each $[b_{ij}] \in (M_n(\mathcal{M})')^+$.

Now if $[a_{ij}] \geq 0$ and $[a_{ji}] \geq 0$, then the fact that
${\mathrm{id}} \otimes \tau$ commutes strongly with $\Delta_n$, surely
ensures that $\Delta_n^{1/4}[a_{ij}]\Omega_n \in \mathcal{P}_n
\cap \mathcal{P}_n^{\tau}$. Moreover for any $[b_{ij}] \in
(M_n(\mathcal{M})')^+$, \cite[Proposition 2.5.26]{BR} alongside the
hypothesis ensures that
$$(T_{\varphi} \otimes
\mathbb{I})^*\Delta_n^{-1/4}[b_{ij}]\Omega_n \in
\overline{co}(\mathcal{P}_n \cup \mathcal{P}_n^{\tau}).$$
In this
case it therefore follows from the duality of $\mathcal{P}_n \cap
\mathcal{P}_n^{\tau}$ and $\overline{co}(\mathcal{P}_n \cup
\mathcal{P}_n^{\tau})$ that $0 \leq
(\Delta_n^{1/4}[a_{ij}]\Omega_n, (T_{\varphi} \otimes
\mathbb{I})^*\Delta_n^{-1/4}[b_{ij}]\Omega_n)$ for each $[b_{ij}]
\in (M_n(\mathcal{M})')^+$, and hence that
$[\varphi(a_{ij})] \geq 0$ as required.

For the converse suppose that
$\overline{\{\Delta_n^{1/4}[a_{ij}]\Omega_n : [a_{ij}] \geq 0,
[a_{ji}] \geq 0\}} = \mathcal{P}_n \cap \mathcal{P}_n^{\tau}$ for
each $1 \leq n \leq k$ and that for each $1 \leq n \leq k$ we have
that $[\varphi(a_{ij})] \geq 0$ whenever $[a_{ij}], [a_{ji}] \in
M_n(\mathcal{M})^+$. To see that then $(T_{\varphi} \otimes
\mathbb{I})^*(\mathcal{P}_n) \subset \overline{co}(\mathcal{P}_n
\cup \mathcal{P}_n^{\tau})$ for each $1 \leq n \leq k$, we need
only show that $(T_{\varphi} \otimes
\mathbb{I})^*(\Delta_n^{-1/4}[b_{ij}]\Omega_n) \subset
\overline{co}(\mathcal{P}_n \cup \mathcal{P}_n^{\tau})$ for each
$1 \leq n \leq k$ and each $[b_{ij}] \in (M_n(\mathcal{M})')^+$
(\cite[Proposition 2.5.26]{BR}). To see that this is indeed the case, the
duality of $\mathcal{P}_n \cap \mathcal{P}_n^{\tau}$ and
$\overline{co}(\mathcal{P}_n \cup \mathcal{P}_n^{\tau})$ ensures
that we need only show that $$0 \leq (\eta, (T_{\varphi} \otimes
\mathbb{I})^*\Delta_n^{-1/4}[b_{ij}]\Omega_n)$$ for each $\eta \in
\mathcal{P}_n \cap \mathcal{P}_n^{\tau}$. In the light of our
assumption regarding $\mathcal{P}_n \cap \mathcal{P}_n^{\tau}$,
this in turn means that we need to show that
\begin{eqnarray*}
0 &\leq& (\Delta_n^{1/4}[a_{ij}]\Omega_n, (T_{\varphi} \otimes
\mathbb{I})^*\Delta_n^{-1/4}[b_{ij}]\Omega_n)\\ &=&
([a_{ij}]\Omega_n, (T_{\varphi} \otimes
\mathbb{I})^*[b_{ij}]\Omega_n)\\ &=& (\varphi_n([a_{ij}])\Omega_n,
[b_{ij}]\Omega_n)
\end{eqnarray*}
for each $[b_{ij}] \in (M_n(\mathcal{M})')^+$ and each $[a_{ij}]
\in M_n(\mathcal{M})$ with $[a_{ij}] \geq 0, [a_{ji}] \geq 0$.
Since by assumption $[\varphi(a_{ij})] \geq 0$ whenever $[a_{ij}]
\geq 0, [a_{ji}] \geq 0$ ($1 \leq n \leq k$), the claim therefore
follows from \cite[Proposition 2.3.19]{BR}.
\end{proof}

\section{Tomita-Takesaki approach for partial transposition}
In order to obtain a more complete characterisation of
$k$-decomposable maps, one should describe elements of the cone
${\mathcal{P}}_k\cap{\mathcal{P}}_k^\tau$ (cf. Theorem \ref{dual}).
In this section we formulate the general scheme for this
description.

Suppose that $A$ is a $C^*$-algebra equipped with a faithful state
$\omega_A$. Let $B=\cB(K_B)$ for some Hilbert space $K_B$, $\varrho$ be
an invertible density matrix in $\cB(K)$ and $\omega_B$ be a state
on $B$ such that $\omega_B(b)=\Tr(b\varrho)$ for $b\in B$. By
$(H,\pi,\Omega)$, $(H_A,\pi_A,\Omega_A)$ and
$(H_B,\pi_B,\Omega_B)$ we denote the GNS representations of
$(A\otimes B, \omega_A\otimes\omega_B)$, $(A,\omega_A)$ and
$(B,\omega_B)$ respectively. We observe that we can make the
following identifications:
\begin{enumerate}
\item $H=H_A\otimes H_B$,
\item $\pi=\pi_A\otimes\pi_B$,
\item $\Omega=\Omega_A\otimes\Omega_B$.
\end{enumerate}
With these identifications we have $\Jm=J_A\otimes J_B$
and $\Delta=\Delta_A\otimes\Delta_B$ where $\Jm$,
$J_A$, $J_B$ are modular conjugations and $\Delta$, $\Delta_A$,
$\Delta_B$ are modular operators for $(\pi(A\otimes B)'',\Omega)$,
$(\pi_A(A)'',\Omega_A)$, $(\pi_B(B)'',\omega_B)$ respectively.
Since $\Omega_A$ and $\Omega_B$ are separating vectors, we will
write $a\Omega_A$ and $b\Omega_B$ instead of $\pi_A(a)\Omega_A$
and $\pi_B(b)\Omega_B$ for $a\in A$ and $b\in B$.

The natural cone $\mathcal{P}$ for
$(\pi(A\otimes B)'',\Omega)$ is defined (see \cite{BR} or \cite{Ara}) as the closure
of the set
$$
\left\{\left(\sum_{k=1}^na_k\otimes b_k\right)j_{\mathrm{m}}\left(\sum_{l=1}^na_l\otimes b_l
\right)\Omega:\,n\in{\mathbb{N}},\,a_1,\ldots,a_n\in A,\,b_1,\ldots,b_n\in B\right\}$$
where $j_{\mathrm{m}}(\cdot)=\Jm\cdot \Jm$ is the modular
morphism on $\pi(A\otimes B)''=\pi_A(A)''\otimes\pi_B(B)''$.

Recall (see Section 4) that $H_B$ is the closure of the set
$\{b\rr:\,b\in B\}$ and $\Omega_B$ can be identified with $\rr$.
Let $U_B$ be the unitary
operator on $H_B$ described in Section 4. Then we have
\begin{lemma}
$({\mathbb{I}}\otimes U_B)\mathcal{P}$ is the closure of the set
$$
\left\{\left(\sum_{k=1}^na_k\otimes\alpha( b_k)\right)j_{\mathrm{m}}\left(
\sum_{l=1}^na_l\otimes\alpha(b_l)
\right)\Omega:\,n\in{\mathbb{N}},\,
\begin{array}{c}a_1,\ldots,a_n\in A\\b_1,\ldots,b_n\in B\end{array}\right\}.
$$ 
\end{lemma}
\begin{proof}
Using the Tomita-Takesaki approach one has
\begin{eqnarray*}
\lefteqn{({\mathbb{I}}\otimes U_B)\left(\sum_k a_k\otimes b_k\right)j_{\mathrm{m}}
\left(\sum_l a_l\otimes b_l\right)\Omega=}\\
&=&\sum_{kl}a_kj_A(a_l)\Omega_A\otimes U_B b_kJ_Bb_lJ_B\Omega_B\\
&=&\sum_{kl}a_kj_A(a_l)\Omega_A\otimes U_B b_k U_BU_BJ_B
b_l\Omega_B\\
&=&\sum_{kl}a_kj_A(a_l)\Omega_A\otimes U_B b_k U_BJ_BU_B
b_l U_BJ_B\Omega_B\\
&=&\left(\sum_k a_k\otimes \alpha(b_k)\right)j_{\mathrm{m}}\left(
\sum_l a_l\otimes\alpha(b_l)\right)
\end{eqnarray*}
In the third equality we used the fact that $U_B$ commutes with $J_B$.
\end{proof}

This leads us to:

\begin{theorem}
Suppose that $K$ is a finite dimensional Hilbert space.
Then $(\jed\otimes U_B){\mathcal{P}}={\mathcal{P}}'$ where
${\mathcal{P}}'$ is the natural cone associated with $(\pi_A(A)\otimes \pi_B(B)',\Omega)$.
\end{theorem}
\begin{proof}
We just proved, that
$(\jed\otimes U_B){\mathcal{P}}$ is the closure of the set
$$
\left\{\left(\sum_{k=1}^na_k\otimes\alpha(b_k)\right)
j_{\mathrm{m}}\left(\sum_{l=1}^na_l\otimes\alpha(b_l)\right)\Omega:
\,n\in{\mathbb{N}},\,a_1,\ldots,a_n\in A,\,b_1,\ldots,b_n\in B\right\}.
$$
By Proposition \ref{przestawianie}(\ref{przestawianie2}) 
$\alpha$ maps $\pi_B(B)''$ onto $\pi_B(B)'$, so
the assertion is obvious.
\end{proof}

Consequently, ${\mathcal{P}}_k\cap{\mathcal{P}}_k^\tau$ is nothing else but
${\mathcal{P}}_k\cap{\mathcal{P}}_k^\prime$.

In the sequel we will assume that $A=\cB(K_A)$ for some finite dimensional Hilbert space
$K_A$ and that $\omega_A$ is determined by some density matrix $\varrho_A$ in $\cB(K_A)$.

\begin{remark}
The operator $\jed\otimes U_B$ is a symmetry in $\cB(H_A\otimes H_B)$ in the sense of \cite{Alf}
(see the paragraph preceding Lemma 6.33).
Obviously, $\jed\otimes U_B$ has a spectral decomposition of the form 
$\jed\otimes U_B=P-Q$ where $P$ and $Q$ are mutually orthogonal projections
in $\cB(H_A\otimes H_B)$ such that $P+Q=\jed$.

Moreover, if $\cS(\cB(H_A\otimes H_B))$ denotes the set of states on $\cB(H_A\otimes H_B)$ 
and $F$ and $G$ are norm closed faces in $\cS(\cB(H_A\otimes H_B))$ associated with
$P$ and $Q$ respectively, then $F$ and $G$ are antipodal and affinely independent faces in
$\cS(\cB(H_A\otimes H_B))$ forming a generalized axis $(F,G)$ of $\cS(\cB(H_A\otimes H_B))$.

Furthermore, the symmetry $\jed\otimes U_B$ provides the one parameter group 
$(\alpha_t^*)_{t\in\bR}$ (where 
$\alpha_t(\cdot)=\exp\left(\frac{it}{2}[\jed\otimes U_B,\cdot]\right)$
for $t\in\bR$) which is the generalised rotation of $\cS(\cB(H_A\otimes H_B))$ 
about $(F,G)$ (cf. \cite[Chapter 6]{Alf}). On the other hand (see again \cite{Alf})
in the algebra $\cB(K_A\otimes K_B)$ there are canonical symmetries associated to
$2\times 2$-matrix units $\{e_{ij}\}$. Moreover these symmetries can be extended 
to a Cartesian triple of symmetries of $\cB(K_A\otimes K_B)$;
a fact which is the basic ingredient 
of the definition of orientation of $\cB(K_A\otimes K_B)$.
By contrast partial transposition yields a symmetry $\jed\otimes U_B$ in the larger algebra
$\cB(H_A\otimes H_B)\supseteq  \cB(K_A\otimes K_B)$ and it would seem that in general
this symmetry  tends to ``spoil"
the orientation sructure of the algebra of interest, i.e. $\cB(K_A\otimes K_B)$.

More precisely: one can repeat the above arguments for $U_B$, so $U_B$ is the symmetry of the 
$\cB(H_B)\supset \cB(K_B)$. The operator $\jed$ is a symmetry of the first factor, being
an element of the smaller algebra $\cB(H_A)\supset \cB(K_A)$. Clearly, this symmetry does not change
the orientation of the algebra of the first factor. As $\jed\otimes U_B$ is the tensor product
of $\jed$ and $U_B$, we ``tranlated" the basic feature of partial transposition --
tensor product of morphism and antimorphism.
\end{remark}

As a clarification of the role of the symmetry $\jed\otimes U_B$ in the structure of 
orientation of $\cB(K_A\otimes K_B)$ is an open question, we wish to collect 
some properties of $\jed\otimes U_B$ in the rest of that 
section.
To this end assume that
$(e_i)$ and $(f_k)$ are othonormal bases in $K_A$ and $K_B$ respectively consisted
of eigenvectors of $\varrho_A$ and $\varrho_B$ respectively;
by $(E_{ij})$ and $(F_{kl})$ we denote matrix units associated with $(e_i)$ and
$(f_k)$ respectively. Each element $a$ of $\cB(K_A)\otimes \cB(K_B)$ can be uniquely written
in the form $a=\sum_{ij}a_{ij}\otimes F_{ij}\equiv[a_{ij}]$ 
for some elements $a_{ij}\in \cB(K_A)$.

Let ${\widetilde U}=\jed\otimes U_B$. Observe that projections $P$ and $Q$ are of the form
$$P=\frac{1}{2}(\jed+ {\widetilde U}),\quad\quad Q=\frac{1}{2}(\jed- {\widetilde U}).$$
At first, we formulate a step towards an
eventual characterisation of the cone $\cP\cap {\widetilde U}\cP$.
\begin{proposition}
\begin{enumerate}
\item
$\cP\cap {\widetilde U}\cP$ is a maximal subcone of $\cP$ which is globally 
invariant with respect to
${\widetilde U}$.
\item
$\cP_A\otimes\cP_B\subset{\widetilde U}\cP$, where $\cP_A \subset H_A$ and
$\cP_B \subset H_B$ are the natural cones respectively
corresponding to the algebras $A$ and $B$ .

\item
Let $U_A$ denote the unitary operator on $H_A$ introduced in section 4 and
$P_A$, $Q_A$ (resp. $P_B$, $Q_B$) be spectral projections of $U_A$ (resp. $U_B$).
For $\xi\in H$ the following are equivalent
\begin{enumerate}
\item $\xi\in\cP\cap {\widetilde U}\cP$; 
\item for every $\eta\in\cP$ we have 
\begin{equation*}
|(\eta,Q\xi)|\leq (\eta,P\xi);
\end{equation*}
\item for every $\eta\in\cP$ we have
\begin{equation*}
(\eta,\xi)\geq 0\quad\mbox{and}\quad 2(\eta,Q\xi)\leq (\eta,\xi);
\end{equation*}
\item
for every $\eta\in\cP$ we have
\begin{eqnarray*}
\lefteqn{(\eta,(P_A\otimes P_B)\xi)+(\eta,(Q_A\otimes P_B)\xi)}\\
&\ge& (\eta,(P_A\otimes Q_B)\xi)+(\eta,(Q_A\otimes Q_B)\xi);
\end{eqnarray*}
\item
for every $\eta\in\cP$ we have
\begin{eqnarray*}
\lefteqn{(\eta,(P_A\otimes P_B)\xi)-(\eta,(Q_A\otimes P_B)\xi)}\\
&\ge& -(\eta,(P_A\otimes Q_B)\xi)+(\eta,(Q_A\otimes Q_B)\xi).
\end{eqnarray*}
\end{enumerate}
\item
If $\xi\in\cP\cap {\widetilde U}\cP$, then $\|Q\xi\|\leq\|P\xi\|$.
\item
$\xi\in\cP\cap {\widetilde U}\cP$ implies that for every $\eta\in\cP$
$$2(\eta, Q_A\otimes Q_B\xi)\leq (\eta, P^{\mathrm{tot}}\xi)$$
where $P^{\mathrm{tot}}=\frac{1}{2}(\jed+U_A\otimes U_B)$.
\end{enumerate}
\end{proposition}
\begin{proof}
Properties (1) and (2) follow from easy observations. In order to prove (3) observe
that both $\xi$ and $ {\widetilde U}\xi$ are in $\cP$, so the selfduality 
of $\cP$ implies that for every $\eta\in\cP$ we have
\begin{eqnarray*}
0&\leq&(\eta,\xi)=(\eta,P\xi)+(\eta,Q\xi),\\
0&\leq&(\eta,{\widetilde U}\xi)=(\eta,P\xi)-(\eta,Q\xi).
\end{eqnarray*}
Thus we have the equivalence of (a) and (b). The equivalence of (a) and (c)
follows from the fact that $\widetilde{U}\xi\in\cP$ is equivalent 
to the following inequality: for every $\eta \in \cP$
$$
0\leq (\eta, {\widetilde U}\xi)=(\eta,P\xi)-(\eta,Q\xi)=(\eta,\xi)-2(\eta,Q\xi).
$$
The rest of (3) can be checked by simple calculations.
To prove (4) assume that
$\xi\in\cP\cap {\widetilde U}\cP$ and $\eta,\eta'\in\cP$. 
From (3) we have
\begin{eqnarray*}
&-(\eta,P\xi)\leq (\eta,Q\xi)\leq (\eta, P\xi)&\\
&-(\eta',P\xi)\leq (-\eta',Q\xi)\leq (\eta', P\xi)&
\end{eqnarray*}
and consequently
$$|(\eta-\eta',Q\xi)|\leq (\eta+\eta',P\xi).$$
To see that (4) holds, we merely need to
apply the above inequality to the case $\eta = \frac{1}{2}\xi$ and
$\eta' = \frac{1}{2}\widetilde{U}\xi$.

It remains to prove (5). Assume $\xi\in\cP\cap \widetilde{U}\cP$.
One can easily check that $(U_A\otimes U_B)\cP=\cP$. Hence, from (3) we have
$((U_A\otimes U_B)\eta,\xi)\geq 0$ and $2((U_A\otimes U_B)\eta,Q\xi)\leq 
((U_A\otimes U_B)\eta,\xi)$. Observe that
\begin{eqnarray*}
((U_A\otimes U_B)\eta,Q\xi)
&=&(\eta,(U_A\otimes U_B)(\jed\otimes Q_B)\xi)=\\
&=&(\eta,(U_A\otimes (P_B-Q_B)Q_B)\xi)=\\
&=&-(\eta,(U_A\otimes Q_B)\xi).
\end{eqnarray*}
Thus we have
\begin{eqnarray*}
-2(\eta,(U_A\otimes Q_B)\xi)&\leq&(\eta,(U_A\otimes U_B)\xi)\\
2(\eta,(\jed\otimes Q_B)\xi)&\leq&(\eta,\xi)
\end{eqnarray*}
where the second inequality follows from (3).
Consequently, we have
\begin{eqnarray*}
2(\eta,(Q_A\otimes Q_B)\xi)&=&
(\eta,(\jed\otimes Q_B)\xi)-(\eta,(U_A\otimes Q_B)\xi)\leq\\
&\leq&\frac{1}{2}[(\eta,\xi)+(\eta,(U_A\otimes U_B)\xi)]=\\
&=&(\eta,P^{\mathrm{tot}}\xi)
\end{eqnarray*}
and the proof is ended.
\end{proof}

Now, assume that $\dim K_A=\dim K_B=2$. If $\xi=\Delta^{1/4}[a_{ij}]\Omega$, 
then $\widetilde{U}\xi=\Delta^{1/4}[a_{ji}]\Omega$, and consequently
\begin{eqnarray*}
P\xi&=&\frac{1}{2}\Delta^{1/4}\left[\begin{array}{cc}
2a_{11}&a_{12}+a_{21}\\a_{12}+a_{21}&2a_{22}\end{array}\right]\Omega,\\[2mm]
Q\xi&=&\frac{1}{2}\Delta^{1/4}\left[\begin{array}{cc}
0&a_{12}-a_{21}\\a_{21}-a_{12}&0\end{array}\right]\Omega.
\end{eqnarray*}
It is easy to observe that if $\xi,\widetilde{U}\xi\in\cP$, then $P\xi\in\cP$. 
Moreover, we have the following
\begin{proposition}
Let $\xi\in\cP$. Then the following are equivalent:
\begin{enumerate}
\item
$Q\xi\in\cP$,
\item
$Q\xi =0$,
\item
$\xi$ is a fixed point of $\widetilde{U}$.
\end{enumerate}
\end{proposition}
\begin{proof}
If $\xi\in\cP$ then $[a_{ij}]$ is positive in $\cB(H_A)$. Then $a_{12}^*=a_{21}$.
Let $b=\frac{1}{2}(a_{12}-a_{12}^*)$. We have that $b=ih$ for some 
selfadjoint element of $\cB(K_A)$ and
$Q\xi=\Delta^{1/4}\left[\begin{array}{cc}0&ih\\-ih&0\end{array}\right]\Omega$.
Now if $P_2\xi\in\cP$, we necessarily have that
$\left[\begin{array}{cc}0&ih\\-ih&0\end{array} \right] \geq 0$.
(See for example the argument used in Lemma \ref{aa}.) However it
is a simple observation that the matrix
$\left[\begin{array}{cc}0&ih\\-ih&0\end{array} \right]$ is
positive if and only if $h=0$, so (1) and (2) are equivalent. The
equivalence of (2) and (3) is evident.
\end{proof}
Hence, in general, $Q\xi$ is not in $\cP$. However, (cf \cite{Ara}), for each $\zeta \in H$
there exists $|\zeta| \in \cP$ such that $\zeta = u |\zeta|$ for some
partial isometry $u$. In the considered case we can calculate $|Q \xi|$
explicitly. Namely we get
\begin{proposition}
Let $ih=v|h|$ be the polar decomposition of element $ih$.
Then 
$Q\xi=\tilde{V}\xi_b$, where 
$\tilde{V}=\Delta^{1/4}\left[\begin{array}{cc}0&v\\-v&0\end{array}\right]\Delta^{-1/4}$ and
$\xi_b\in\cP$.
\end{proposition}
\begin{proof}
Let $B=\left[\begin{array}{cc}0&ih\\-ih&0\end{array}\right]$ and $V=\left[\begin{array}{cc}
0&v\\-v&0\end{array}\right]$.
Then one can check that
$B=V\left[\begin{array}{cc}
|h|&0\\0&|h|\end{array}\right]$
is the polar decomposition of $B$.
Furthermore we have
$$Q\xi=\Delta^{1/4}B\Omega=\Delta^{1/4}V|B|\Omega=\Delta^{1/4}V\Delta^{-1/4}
\Delta^{1/4}|B|\Omega=\tilde{V}\xi_b$$
where $\xi_b=\Delta^{1/4}|B|\Omega$ is an element of $\cP$.
\end{proof}

Here is another way of writing $Q\xi$. Namely, 
there is $|Q\xi|\in\cP$ such that $Q\xi=u|P_2\xi|$ where
$u$ is a partial isometry such that $u\in (A\otimes B)'$, $uu^*=[(A\otimes B)'Q\xi]$
and $u^*u=[(A\otimes B)'|Q\xi|]$ (cf. \cite{Ara}). Moreover, one can check that
$$
Q\xi=\left[\begin{array}{cc}\alpha&0\\0&-\alpha^{-1}\end{array}\right]
\left[\begin{array}{cc}
0&\varrho_A^{1/4}b\varrho_A^{-1/4}\\
-\varrho_A^{1/4}b\varrho_A^{-1/4}&0\end{array}\right]$$
where $\alpha=\lambda_1^{1/4}\lambda_2^{-1/4}$ and $\lambda_1,\lambda_2$ are eigenvalues
of $\varrho_B$ ($b$ was defined in the proof of Proposition 6.5).

What is still lacking is an explicit description of the role
of the symmetry $\widetilde{U}$ in terms of the algebra $\cB(K_A) \otimes \cB(K_B)$.
This will be done in the forthcoming paper \cite{LMM}.

\bibliographystyle{amsplain}

\end{document}